\documentclass[aps,prb,onecolumn,10pt,groupedaddress,nofootinbib]{revtex4}
\usepackage[utf8x]{inputenc}
\usepackage{amsmath,amssymb}
\usepackage[dvips]{graphicx}
\begin{document}

\title{Comment on ``Critical point scaling of Ising spin glasses in a magnetic field'' by J.~Yeo and M.A.~Moore}

\author{T.~Temesv\'ari}\email{temtam@helios.elte.hu}
\affiliation{MTA-ELTE Theoretical Physics Research Group,
E\"otv\"os University, P\'azm\'any P\'eter s\'et\'any 1/A,
H-1117 Budapest, Hungary
}

\date{\today}

\begin{abstract}
 In a section of a recent publication, [J.\ Yeo and M.A.\ Moore, Phys.\ Rev.\ B {\bfseries 91}, 104432 (2015)], the authors
 discuss some of the arguments in the paper by Parisi and Temesv\'ari [Nuclear Physics B {\bfseries 858}, 293 (2012)].
 In this comment, it is shown how these arguments are misinterpreted, and the existence of the Almeida-Thouless transition
 {\em in\/} the upper critical dimension 6 reasserted.
\end{abstract} 


\maketitle

In a recent paper \cite{Yeo_Moore_2015} by Yeo and Moore about the long debated existence of the Almeida-Thouless instability \cite{AT}
in the short ranged Ising spin glass below the upper
critical dimension six, the authors criticize in Sec.\ III.\ some of our statements and arguments in Ref.\ \cite{PT}. In that paper we have
demonstrated: firstly the incorrect reasoning of Ref.\ \cite{Moore_Bray_2011} about the disappearence of the Almeida-Thouless (AT) transition line when approaching the upper
critical dimension from above; secondly we have computed the AT line staying exactly in six dimensions (and not by a limiting process); and thirdly the $\epsilon$-expansion
was used to compute the AT line below six dimensions, and the relatively smooth behavior of it while crossing $d=6$ (with fixed bare parameters) was exhibited. 
In what follows,
we want to comment the discussion in Sec.\ III.\ of \cite{Yeo_Moore_2015}.

\section{At and above six dimensions}
\label{1}

The first order renormalization group (RG) equations for the six-dimensional model are worked out and solved in Sec.\ 3 of Ref.\ \cite{PT}, the AT line follows from that
calculation [see Eq.\ (37) in \cite{PT}]\footnote{We use here the notations of Ref.\ \cite{Yeo_Moore_2015}. In fact $|r|$ was called $\tau$ in \cite{PT}, whereas $r$
had the role of the nonlinear scaling field associated with $\tau$. We also adapt here to the somewhat unconventional use of the symbol $\epsilon$ as
$\epsilon=d-6$.}:
\begin{equation}\label{AT_d=6}
h^2_{\text{AT}}=\dfrac{4}{(1-w^2\ln |r|+\frac{10}{3}w^2\ln w)^4}\,\, w|r|^2 \approx \dfrac{4}{(1-w^2\ln |r|)^4}\,\, w|r|^2, \qquad\qquad d=6
\end{equation}
where $w^2\ll 1$ was used. [Note that a minus sign in the denominator of Eq.\ (13) has been left out in \cite{Yeo_Moore_2015}.]
As it turns out from the discussion in Sec.\ 3 of Ref.\ \cite{PT}, this approximation is valid if the scaling variable with zero scaling dimension
(which is invariant under RG in $d=6$) is small, i.e.
\begin{equation}\label{d=6_inequality}
\dfrac{w^2}{1+\frac{5}{3}w^2\ln w^2-w^2\ln |r|}\ll 1\,,
\end{equation}
and this condition is always satisfied whenever $|r|\ll 1$ and $w^2\ll 1$; see also the middle part of Eq.\ (59) of that reference.
Yeo and Moore \cite{Yeo_Moore_2015} forget all about this derivation of the six-dimensional AT line; they deduce it from Eq.\ (11) of \cite{Yeo_Moore_2015}
by the limit $\epsilon\to 0$, and finally
they argue that ``Eq.\ (11) is not valid for this limit''. We can absolutely agree with this last statement: the system at the upper critical dimension needs special care,
physical quantities, like the critical magnetic field where replica symmetry breaking sets in, cannot be obtained by a limiting process of $\epsilon\to 0$. The point is that
$\epsilon$ in Eq.\ (11) may be small, but fixed, while $|r|\ll 1$, and the $|r|^{\epsilon/2}$ term in the denominator must be ignored. Taking account of this, the AT line
above dimension six, Eq.\ (11) of \cite{Yeo_Moore_2015}, must be written (consistently with the approximations used to derive it) as:
\begin{equation}\label{AT_d>6}
h^2_{\text{AT}}\sim \dfrac{w|r|^{\frac{d}{2}-1}}{(\frac{2w^2}{\epsilon}+1)^{\frac{5d}{6}-1}}\,,  \qquad\qquad d>6.
\end{equation}
This is just Eq.\ (28) of Ref.\ \cite{PT}.
This equation for the AT line above six dimensions must be supplemented by the range of its applicability, otherwise false conclusions like Eq.\ (12) in \cite{Yeo_Moore_2015}
[which is obviously
incompatible with (\ref{AT_d=6})] could be deduced. For this reason, we briefly repeat the two steps needed for the derivation of (\ref{AT_d>6}):
\begin{itemize}
 \item 
 The RG equations for the three bare parameters, namely 
 \begin{equation}\label{RG_d>6}
 \begin{aligned}
  \dot{|r|}&=\left(2-\frac{10}{3}w^2\right)\,|r|,\\  
  \dot{w^2}&=-\epsilon w^2-2w^4,\\
  \dot{h^2}&=\left(4+\frac{\epsilon}{2}+\frac{1}{3}w^2\right)\,h^2
 \end{aligned}
 \end{equation}
are valid for $|r|\ll 1$ and $w^2\ll 1$. One can introduce the nonlinear scaling fields \cite{Wegner} satisfying exactly, by definition, the linearized
(around the fixed point) and diagonalized RG equations. For the system in (\ref{RG_d>6}), and for its Gaussian fixed point, one readily finds
\[
\dot{g_{|r|}}=2\,g_{|r|},\qquad \qquad    \dot{g_{w^2}}=-\epsilon\, g_{w^2}, \qquad \text{and} \qquad \dot{g_{h^2}}=\left(4+\frac{\epsilon}{2}\right)\,g_{h^2}.
\]
The relations between bare parameters and nonlinear scaling fields were published in \cite{PT}, for completeness we repeat them here:
\begin{equation}\label{bare_vs_g}
|r|=g_{|r|}\,\left(1-\frac{2}{\epsilon}\,g_{w^2}\right)^{-\frac{5}{3}},\quad w^2= g_{w^2}\,\left(1-\frac{2}{\epsilon}\,g_{w^2}\right)^{-1},\quad 
\text{and} \quad h^2= g_{h^2}\,\left(1-\frac{2}{\epsilon}\,g_{w^2}\right)^{\frac{1}{6}}.
\end{equation}

\item
The zeros of the scaling function of the replicon mass, $\hat \Gamma_R$, are the locations of the AT instability. $\hat \Gamma_R$ depends on the bare parameters
$|r|$, $w^2$, and $h^2$ through the RG invariants $x\equiv g_{w^2}\,g_{|r|}^{\frac{\epsilon}{2}}$ and $y\equiv g_{h^2}\,g_{|r|}^{-2-\frac{\epsilon}{4}}$.
The AT instability line can then be written as $y=f(x)$ or 
\begin{equation}\label{generic_AT}
 g_{h_{\text{AT}}^2}=g_{|r|}^{2+\frac{\epsilon}{4}}\,f\left(g_{w^2}\,g_{|r|}^{\frac{\epsilon}{2}}\right)
 =\dfrac{g_{|r|}^{2}}{\sqrt{g_{w^2}}}\,g\left(g_{w^2}\,g_{|r|}^{\frac{\epsilon}{2}}\right),
 \qquad \text{with}\qquad g(x)\equiv \sqrt{x}\,f(x).
\end{equation}

The following remarks are now in order:
\begin{description}
 \item[(i)]  This form of the AT line is generic for the system where the zero-external-magnetic-field symmetry is broken only
by the linear replica symmetric invariant in the Lagrangian whose bare coupling constant is $h^2$.
(This model is used in Refs.\cite{Moore_Bray_2011,Yeo_Moore_2015} too.) Eqs.\ (\ref{bare_vs_g}) cannot be used, in this generic case, to replace
nonlinear scaling fields by bare couplings, as they were derived from the one-loop RG equations in (\ref{RG_d>6}).
 \item[(ii)] Eq.\ (14) of \cite{Yeo_Moore_2015} formally agrees with (\ref{generic_AT}), but the bare couplings are there instead of the $g$'s.
In this form it is not correct.
\item[(iii)] The function $g(x)$ of (\ref{generic_AT}) can be calculated perturbatively, the 1-loop result was published in \cite{PT}: $g(x)=(-C')\,x$
where $-C'(\epsilon)>0$ is analytic and positive around $\epsilon=0$. Putting this into (\ref{generic_AT}), one gets
\[
 g_{h_{\text{AT}}^2}\sim g_{|r|}^{2+\frac{\epsilon}{2}}\,\sqrt{g_{w^2}}\quad,
 \]
and inserting the inverse relations of those in Eq.\ (\ref{bare_vs_g}) one immediately arrives at (\ref{AT_d>6}).
\end{description}
\end{itemize}
As it must be clear from the two-step process above, a mixture of renormalization {\em and\/} perturbation theory leads to Eq.\ (\ref{AT_d>6}).
The leading, linear contribution to $g(x)$  is free from a singularity at $d=6$, as it comes from an ultraviolet convergent one-loop graph \cite{PT}.
Triangular insertions in the next, two-loop graphs, however, certainly produce singular terms like $g(x)\sim \frac{1}{\epsilon}\,x^2$, their neglect
is acceptable only if $\frac{1}{\epsilon}\,x=\frac{1}{\epsilon}\,g_{w^2}\,g_{|r|}^{\frac{\epsilon}{2}}\ll 1$. Expressing this condition by the 
bare couplings, one can write the range of applicability of Eq.\ (\ref{AT_d>6}) as
\begin{equation}\label{d>6_inequality}
|r|\ll 1, \qquad w^2\ll 1, \qquad \text{and most importantly} \qquad \dfrac{1}{\epsilon}\,w^2\,|r|^{\frac{\epsilon}{2}}\,
\left(1+\frac{2}{\epsilon}\,w^2\right)^{-1-\frac{5}{6}\epsilon}\ll 1.
\end{equation}
The left-hand-side of the third condition becomes of order unity ($1/2$), and thus breaks down, when $\epsilon\to 0$ while $|r|$ and $w^2\ll 1$,
but otherwise fixed. This is just the limit leading to Eq.\ (12) of \cite{Yeo_Moore_2015} (and to the conclusion of the disappearance of the AT line
for $\epsilon\to 0$), and is the source of the basic fault in the original arguments in \cite{Moore_Bray_2011}.
[See also Fig.~2(b) and the discussion around it
in \cite{PT}.] $\epsilon$ in (\ref{AT_d>6}) may be small, but must be kept fixed. Simple first order perturbational result is obtained for
$w^2\ll\epsilon$. The joint application of the perturbational method and RG (and not RG alone as Yeo and Moore \cite{Yeo_Moore_2015} claim)
provide (\ref{AT_d>6}) which is valid for $0<\epsilon\ll w^2\ll 1$ too.
In this latter case the range of applicability of Eq.\ (\ref{AT_d>6}), according to (\ref{d>6_inequality}), shrinks to zero as $-\ln|r|\gg \epsilon^{-1}$,
together with the amplitude in (\ref{AT_d>6}). This phenomenon signals the appearance of the logarithmic correction in $d=6$:
$h^2_{\text{AT}}\sim (\ln|r|)^{-4}\,|r|^2$, and it is not an indication of the disappearance of the AT line.

\end{document}